\documentclass[a4paper,10pt]{article}

\usepackage{cite}
\usepackage{color}
\usepackage{graphicx}
\usepackage[usenames,dvipsnames]{xcolor} 

\definecolor{DarkGreen}{RGB}{0,64,0}

\usepackage[]{array}
\usepackage[]{amsmath}
\usepackage[]{amssymb}
\usepackage[]{mathrsfs}
\usepackage[]{tensor}
\usepackage{bm}

\begin{document}

\vspace*{2cm}

\begin{center}

{\LARGE{\bf Curing black hole singularities with local scale\\[2mm] invariance}}
\vskip 1cm

{\Large Predrag Dominis Prester}\\
{}~\\
\quad \\
{\em Department of Physics, University of Rijeka,}\\[1mm]
{\em  Radmile Matej\v{c}i\'{c} 2, HR-51000 Rijeka, Croatia}\\
\vskip 1cm
Email: pprester@phy.uniri.hr

\end{center}

\vskip 1.5cm 

\noindent
{\bf Abstract.} 

\medskip

\noindent
We show that Weyl-invariant dilaton gravity provides a description of black holes without classical spacetime singularities. Singularities appear due to ill-behaviour of gauge fixing conditions, one example being the gauge in which theory is classically equivalent to standard General Relativity. The main conclusions of our analysis are: (1) singularities signal a phase transition from broken to unbroken phase of Weyl symmetry, (2) instead of a singularity there is a ``baby-universe'' or a white hole inside a black hole, (3) in  the baby universe scenario there is a critical mass after which reducing mass makes black hole larger as viewed by outside observers, (4) if a black hole could be connected with white hole through the ``singularity'', this would require breakdown of (classical) geometric description, (5) the singularity of Schwarzschild BH solution is non-generic and so it is dangerous to rely on it in deriving general results. Our results may have important consequences for resolving issues related to information-loss puzzle. The theory we use is basically a completion of General Relativity, containing neither additional physical excitations nor higher-derivative terms, but requires physical scalar field such as Higgs field of Standard Model. Though quantum effects are still crucial and may change the proposed classical picture, a position of building quantum theory around essentially regular classical solutions normally provides a much better starting point.

\vskip 1cm 

 
\vfill\eject

\section{Introduction}
\label{sec:intro}


Black hole (BH) solutions in General Relativity (GR) typically contain space-time singularities, i.e., hypersurfaces which particles or observers may hit in finite proper time, but on which curvature blows up to infinity. A textbook example is Schwarzschild black hole
\begin{equation} \label{SchBH}
ds^2 = -\left(1-\frac{2GM}{r}\right) dt^2 + \left(1-\frac{2GM}{r}\right)^{\!-1} dr^2 
+ r^2 (d\theta^2 + \sin^2\theta d\varphi^2) \;,
\end{equation}
where $G$ is Newton constant and $M$ is the mass of the BH. A singularity is located at $r=0$, which can be established by calculating Kretschmann invariant $R_{\mu\nu\rho\sigma} R^{\mu\nu\rho\sigma} \sim r^{-6}$, and showing that all objects entering BH interior ($r < 2GM$) end there in finite proper time. The general problem is that classical physical description completely breaks-down on such space-time singularities, and objects hitting them simply cease to exist (after being smashed by gravity). Unfortunately, there is a large body of theoretical analyses strongly indicating that singularities inside BH's, as well as cosmological ones (Big Bang being an example), are generically unavoidable in the classical GR. 

As there is a neighbourhood around singularity where space-time curvature radii become smaller then Planck length, the usual philosophy is to seek a rescue in the quantum gravity. However, standard physical reasoning would prefer situation where classical singularities are not present, even if quantum description is important.  Our goal here is to argue that one can achieve this in a rather conservative way without introducing new degrees of freedom, higher derivative or non-local terms in the action.\footnote{That some higher-derivative theories may resolve singularities is observed before (e.g., see \cite{Kanti:1998jd,Modesto:2011kw,Biswas:2011ar}). However, such theories either include new degrees of freedom, have problems with unitarity, or are non-local.} In essence, one just needs to redefine the set of fundamental fields. For the reasons of simplicity and clarity, we shall restrict our attention here to neutral spherically symmetric BHs, but we shall argue that the main conclusions are valid more generally.

The outline of the paper is as follows. In Sec.\ \ref{sec:wysm} we set a stage with a brief review of Weyl-invariant dilaton gravity. In Sec.\ \ref{sec:example} we take one particular example of BH solution on which we explicitly demonstrate our main results. In Sec.\ \ref{sec:general} we argue that these result are valid for generic spherically symmetric BHs, and possibly for all generic BHs which inclose a spacelike singularity in GR description. We show that singularity of Schwarzschild BH solution is non-generic and probably non-physical. In Sec.\ \ref{sec:concl} we conclude by summarising our main results and commenting possible relevancy for information loss puzzle and cosmological singularities.


\section{Weyl-invariant dilaton gravity}
\label{sec:wysm}


Let us assume that the (physical) field content of a theory consists of the matter sector $\psi = \{\psi_i\}$ (all fields with nonzero spin except the graviton) and one scalar spin-0 field $s$, and that the matter part of the action $\mathcal{A}_{\mathrm{matt}} = \int dx^4 \sqrt{-g}\, \mathcal{L}_{\mathrm{matt}}(s,\psi)$ is locally scale invariant (Weyl-invariant). The Standard Model of particle physics with minimal Higgs sector is such a theory. Then the simplest way to introduce dynamics to gravity by making the full classical action Weyl-invariant is to take Lagrangian to be
\begin{equation} \label{Lwism}
\mathcal{L}_{\mathrm{WIDG}} = \frac{1}{12} (\phi^2 - s^2) R 
 + \frac{1}{2} (\partial \phi)^2 - \frac{1}{2} (\partial s)^2 - V(\phi,s)
 + \mathcal{L}_{\mathrm{matt}}(s,\psi)
\end{equation}
where $\phi$ is the new scalar field (``dilaton''). The scalar potential is restricted to the form
\begin{equation} \label{Vpot}
V(\phi,s) = \phi^4 f(s/\phi)
\end{equation}
where $f$ is an arbitrary function of the gauge invariant $s/\phi$. The action then is invariant on Weyl transformations defined by
\begin{eqnarray}
g^{\mu\nu}(x) \to \Omega(x)^2\, g^{\mu\nu}(x) ,\quad \phi(x) \to \Omega(x)\, \phi(x) ,\quad 
s(x) \to \Omega(x)\, s(x) , \quad \psi_i(x) \to \Omega(x)^{d_i}\, \psi_i(x)
\label{weyltr}
\end{eqnarray}
where $\Omega(x) \ne 0$ is otherwise arbitrary smooth function and $\{d_i\}$ are canonical scaling dimensions of the matter fields $\{\psi_i\}$. Such models are sometimes called Weyl-invariant dilaton gravity (WIDG). \footnote{In principle one could add higher-curvature terms in the action, e.g., at four-derivatives one allowed term is $\Delta\mathcal{L}_C = h(s/\phi)\, C^2$ where $C_{\mu\nu\rho\sigma}$ is the Weyl tensor and $h$ is an arbitrary function. As we want to stick to two-derivate theories we shall not include such terms here.} The formalism can be extended to include several scalar fields (e.g., describing non-minimal Higgs sector, inflaton, axion) with possibly non-canonical kinetic terms 
\cite{Bars:2013yba} (and, if one allows direct coupling of the dilaton $\phi$ to the matter sector, to any 
$\mathcal{L}_{\mathrm{matt}}$). Though the general idea is rather old and goes back at least to \cite{Dirac:1973gk}), only recently WIDG theory (\ref{Lwism}) received serious attention from a phenomenological viewpoint, in particular in cosmological settings (see, e.g., \cite{Bars:2013yba,Kallosh:2013hoa}) and in studies of a high-energy behaviour of the (canonical) quantum gravity \cite{Hooft:2010ac,tHooft:2011aa}.

With Weyl-invariance an additional gauge symmetry is introduced. Though we have the additional field (dilaton $\phi(x)$), by using the gauge freedom (\ref{weyltr}) we can gauge-fix one field component so the total number of physical degrees of freedom is as in GR. In ``normal'' circumstances $\phi$ and $g^{\mu\nu}$ are nonvanishing thus causing Weyl symmetry to be spontaneously broken.

WIDG is not some exotic way of incorporating gravity into the theory. This is best seen in a so-called E-gauge which is defined by the following gauge-fixing condition
\begin{equation} \label{Egauge}
(\phi_E^2 - s_E^2) = \mathrm{const} = \frac{3}{4\pi G_N}
\end{equation}
This allows one to use the following parametrisation
\begin{equation} \label{phigE}
\phi_{\mathrm{E}}(x) = \pm \sqrt{\frac{3}{4\pi G_N}}\, \cosh \left( \sqrt{\frac{4\pi G_N}{3}} \sigma(x) \right)
\qquad, \qquad
s_{\mathrm{E}}(x) = \pm \sqrt{\frac{3}{4\pi G_N}}\, \sinh \left( \sqrt{\frac{4\pi G_N}{3}} \sigma(x) \right)
\end{equation}
so that in E-gauge WIDG Lagrangian becomes
\begin{equation} \label{LwismE}
\mathcal{L}^{({\mathrm{E}})}_{\mathrm{WIDG}} = \frac{R}{16\pi G_N} - \frac{1}{2} (\partial \sigma)^2
 - V_{\mathrm{E}}(\sigma) + \mathcal{L}_{\mathrm{matt}}(\sigma,\psi)
\end{equation}
where $V_{\mathrm{E}}(\sigma) = V(\phi_{\mathrm{E}},s_{\mathrm{E}})$. We see that in the domain where E-gauge is applicable WIDG is classically equivalent to the standard GR implementation of gravity, with $G$ playing the role of Newton constant. However, as $s/\phi$ is a gauge invariant, we see that E-gauge (\ref{Egauge}) (with finite $G$) can be defined only in the part of the configuration space satisfying $|\phi| > |s|$ with all configurations finitely separated from $|\phi| = |s|$ in E-gauge. One may say that WIDG is not an alternative theory to GR but its (Weyl-invariant) completion
\cite{Bars:2013yba}.

Our philosophy here is to take $\{\phi,s,g^{\mu\nu},\psi\}$ as the fundamental fields. This means that for physically acceptable solutions these fields should be well-behaved, at least in a class of gauges which we shall call regular gauges. Note that the gauge-invariant scalar field $\sigma$, which from (\ref{phigE}) is defined by
\begin{equation} \label{sigdef}
\sigma = \sqrt{\frac{3}{4\pi G_N}} \, \tanh^{-1} \left( \frac{s}{\phi} \right) 
\end{equation}
can be singular even when $\phi$ and $s$ are perfectly regular -- this happens at the points where $|s/\phi| \to 1$. Recently, in the series of papers \cite{Bars:2011mh,Bars:2011aa,Bars:2012mt} it was demonstrated on explicit examples how cosmological singularities present in GR appear when $|s/\phi| \to 1$ and so are a consequence of a singular behaviour of E-gauge. Motivated by this result, we set as our goal here to investigate the nature of BH singularities in WIDG description.


\section{An example}
\label{sec:example}


Before discussing general BHs it is good to have some explicit example for a demonstration. This section serves the purpose, and later on we shall argue that the main results are generic in the WIDG description of BHs. Now, for practical purposes we need analytic solution, but with the nontrivial (i.e., not constant) scalar field because generic (dynamical) BH configurations are such. So, we can take one of the "hairy" spherically symmetric BH solutions from the literature, where the simplest contain just the metric and one physical scalar field as degrees of freedom. The simplest candidate appears to be BBM solution \cite{BBM}, but it was shown in \cite{Prester:2014voa} that this solution does not describe proper BH in the WIDG theory. The same applies to various generalisations of BBM solution, such as those described in \cite{Bekenstein:1974sf,Martinez:2002ru,Cebeci:2000pt}.\footnote{As shown in \cite{Prester:2014voa}, from the viewpoint of  WIDG theory these solutions describe "collections" of spacetimes without an event horizon, such as wormholes and naked singularities.}

One such simple analytic example of "hairy" BH solution is Zloshchastiev solution \cite{Zloshchastiev:2004ny}.\footnote{We could instead use any of the proper "hairy" BH solutions from \cite{Anabalon:2013qua,Anabalon:2012dw}. The results would be essentially the same.} In E-gauge the Lagrangian (\ref{LwismE}) is specified by 
$\mathcal{L}_{\mathrm{matt}} = 0$ and the scalar potential
\begin{equation} \label{Zpot}
V_{\mathrm{E}}(\sigma) = 4\lambda [3 \sinh\sigma - \sigma (\cosh\sigma + 2)]
\end{equation}
where $\lambda$ is a coupling constant. The potential is obviously unbounded from below, which could make one uncomfortable. However, we shall argue in Sec. \ref{sec:general} that this does not corrupt the analysis, and that conclusions obtained below can be obtained from all known spherically symmetric proper BH solutions with nontrivial scalar field, also in physically acceptable theories, embedded in WIDG formalism. 

This theory has two types of static spherically symmetric asymptotically flat BH solutions. Beside the standard Schwarzschild BH solution, for which space-time metric is as in (\ref{SchBH}) and $\sigma(x) = 0$, there is also another "hairy" branch given by \cite{Zloshchastiev:2004ny}
\begin{equation} \label{Zmetric}
ds_{\mathrm{E}}^2 = - N(r)\, dt^2 + \frac{dr^2}{N(r)} + R(r)^2\, (d\theta^2 + \sin^2\theta d\varphi^2) \;, \;, \qquad
\sigma = \ln \left( 1 + \frac{\kappa}{r} \right)
\end{equation}
where
\begin{equation} \label{Zsoldef}
N(r) = 1 - \lambda \left[ \kappa (2 r + \kappa) - 2 R(r)^2 \ln\! \left( 1 + \frac{\kappa}{r} \right) \right]
\;,\qquad
R(r) = \sqrt{r(r+\kappa)}
\end{equation}
The integration constant $\kappa$ is connected with ADM mass $M$ through $M=\lambda\kappa^3/(6G_N)$. We have assumed $\lambda>0$ (there is \{$\sigma \to -\sigma$, $\lambda \to -\lambda$\} symmetry), and we used the convention $16\pi G_N = 1$.

The main properties and the thermodynamical behaviour of this BH are qualitatively similar to those of Schwarzschild BH. In E-gauge there is a singularity at $r=0$, signalled by behaviour of curvature invariants $(R_{(\mathrm{E})\mu\nu\rho\sigma})^2$, $(R_{(\mathrm{E})\mu\nu})^2$, $R_{(\mathrm{E})}^2$ which all diverge as $\sim r^{-4}$ near $r=0$. The scalar field also diverges at the singularity as $\sigma \to -\ln r \to \infty$, which by (\ref{phigE}) implies $\phi_{\mathrm{E}} \to \infty$ and $s_{\mathrm{E}} \to \infty$. For $\kappa > \lambda^{-1/2}$ there is a regular event horizon, located at $r = r_h > 0$ with $r_h$ defined implicitly by $N(r_h) = 0$, which hides singularity from the outer world in accord with Cosmic Censorship Conjecture. An object which enters BH interior $r<r_h$ unavoidably reaches $r=0$ in a finite proper time, so the space-time is geodesically incomplete. Larger BH (larger proper horizon area $A_{(\mathrm{E})h} = 4\pi r_h(r_h+\kappa)$) means larger entropy $S = A_{(\mathrm{E})h}/(4G_N)$), larger mass $M$, but smaller BH temperature. Normal BH (with $\kappa > \lambda^{-1/2}$) by emitting Hawking radiation is expected to shrink (in the classical analysis) to $r_h \to 0$, $A_{\mathrm{E}h} \to 0$ and to finite mass 
$M \to 1/(6G_N \lambda^{1/2})$. When BH size becomes of the order of the Planck length ($r_h \sim 1$), one expects quantum gravity to be strong and to dictate the final outcome of the process. In the rest of this section we focus on the BH (\ref{Zmetric})-(\ref{Zsoldef}), and postpone the analysis of Schwarzschild BH to Sec. \ref{sec:general}.\footnote{In \cite{Zloshchastiev:2004ny} it is argued that the above model, or its generalisation which includes cosmological constant, may be interesting from the cosmological perspective. However, we emphasise again that the sole purpose of this BH solution here is to provide us with an analytic nontrivial toy model which we use to build general arguments. Those who want more details about Zloshchastiev BH solution, properties of the potential (\ref{Zpot}) and its possible relevance in cosmology should consult Ref.\ \cite{Zloshchastiev:2004ny}.}

We now proceed to show that in WIDG picture $r=0$ is not a space-time singularity but gauge artefact, by constructing gauges in which the solution is perfectly regular for $r>0$. First, we observe that $\sigma \to \infty$ for $r \to 0$, which when used in (\ref{sigdef}) gives $s/\phi \to 1$. This is a gauge-invariant result. Using this in (\ref{Egauge}) immediately signals that E-gauge breaks at $r=0$, so we need to find some non-singular gauge. To keep manifest diff-covariance, and also avoid possible higher-derivative field transformations, let us restrict here to gauge-fixing conditions of the form 
$\mathcal{G}(\phi,s) = 0$. We now show that gauges, which we call $p$-gauges, labeled by the parameter $p > 1$ and defined by the condition
\begin{equation} \label{pgauge}
\frac{1}{12} (\phi_p(x)^2 - s_p(x)^2) = \exp(-2p\,\sigma(x)) \;, \qquad p > 1 \;,
\end{equation}
do the job. The gauge-invariant scalar field $\sigma(x)$ is defined in (\ref{sigdef}). From (\ref{Egauge}) and (\ref{pgauge}) follows that the transition from E-gauge to $p$-gauge is accomplished with the scaling factor $\Omega_p(x) =  \exp(-p\,\sigma(x))$. Using this we can immediately write the solution (\ref{Zmetric})-(\ref{Zsoldef}) in the $p$-gauge, which is
\begin{equation} \label{Psol}
g_{(p)\mu\nu} = \left( 1 + \frac{\kappa}{r} \right)^{\!2p} g_{(\mathrm{E})\mu\nu}
\;, \qquad
\left\{ \begin{matrix} \phi_p(x) \\ s_p(x) \end{matrix} \right\}
 = \sqrt{12} \left( 1 + \frac{\kappa}{r} \right)^{\!-p} \left\{ \begin{matrix} \cosh \\ \sinh \end{matrix} \right\}
 \left( \frac{1}{\sqrt{12}} \ln \left( 1 + \frac{\kappa}{r} \right) \right)
\end{equation}
where $g_{(\mathrm{E})\mu\nu}$ is given in (\ref{Zmetric})-(\ref{Zsoldef}). 

\medskip
\noindent
{\bf (a) Regular solution without singularities for $r>0$.}\\[1mm]
It is obvious that (\ref{Psol}) describes asymptotically flat static spherically symmetric BH solution with event horizon located at the same $r_h$, obtained from $N(r_h) = 0$. The main difference, compared to E-gauge result, is behaviour near $r=0$. Note first that scalar fields $\phi_p$ and $s_p$ are well-defined and vanishing in the limit $r\to0$. Quadratic curvature invariants $R_{(p)}^2$, $(R_{(p)\mu\nu})^2$ and $(R_{(p)\mu\nu\rho\sigma})^2$ all behave as $\sim r^{4(p-1)}$ near $r=0$. In fact, it can be shown that any curvature invariant constructed by tensorial multiplication of $n$ Riemann tensors and $m$ covariant derivatives evaluated on metric (\ref{Psol}) behaves like $\sim r^{(2n+m)(p-1)}$ when $r\to0$, so it is regular and vanishing for $p>1$. This already suggests that $r=0$ is not a singularity in $p$-gauges with $p>1$. To further understand properties of $r=0$ surface, it is necessary to analyse proper distances and particle trajectories near $r=0$. For 
$r \ll \kappa$ the metric (\ref{Psol}) is approximately given by 
\begin{equation} \label{Zpgr0}
ds_p^2 \approx \left(\frac{\kappa}{r} \right)^{\!2p}
 \left[ (\lambda \kappa^2 - 1) dt^2 - \frac{dr^2}{\lambda \kappa^2 - 1} + \kappa\, r (d\theta^2 + \sin^2\!\theta\, d\varphi^2) \right] \;, \qquad r \ll \kappa
\end{equation}
We note that in the ``static" (or Schwarzschild-like) coordinate system we employ (which is singular at the horizon), roles of $t$ and $r$ coordinates interchange for $r < r_h$, so $r$ variable becomes ``time" coordinate, and $t$ spacial coordinate. From (\ref{Zpgr0}) we obtain that the proper time separation between two points with the same spatial coordinates, 
$(r_0,t_0,\theta_0,\varphi_0)$ and $(r,t_0,\theta_0,\varphi_0)$, behaves for $r < r_0 \ll \kappa$ as 
\begin{equation} \label{prosep}
\Delta \tau_{p} \approx \frac{\kappa^p}{\sqrt{\lambda \kappa^2 - 1}} 
\left( r^{1-p} - r_0^{1-p} \right)
\end{equation}
We see that timelike hypersurface $r=0$ is at an infinite proper distance from any point with fixed $r>0$. Similarly, for radial geodesics the proper time also behaves as $\tau \sim r^{1-p}$. 

However, classical trajectories of particles interacting with fields $\phi$ or $s$ are not described by geodesics, but by 
Weyl-invariant action
\begin{equation} \label{Aparticle}
\mathcal{A}_{\mathrm{part}} = - \int \phi(x)\, \chi(s(x)/\phi(x))
 \sqrt{- g_{\mu\nu}(x)\, dx^\mu\, dx^\nu} + \ldots
\end{equation}
where ``$\ldots$'' stands for terms describing interaction with background fields from the matter sector which we neglect here. As for $r\ll\kappa$ we have $s/\phi \approx 1$, the simplest choice is to assume 
$\chi(s/\phi) \approx \chi_0 = \mathrm{const.}$ in (\ref{Aparticle}) (though we shall later entertain other possibilities). Then one gets the following equations of motion for particle trajectories \cite{Bekenstein:1975ts}
\begin{equation} \label{wgpe}
\frac{d^2 x^\mu}{d\lambda^2} + \Gamma^\mu_{\nu\rho} \frac{dx^\nu}{d\lambda} \frac{dx^\rho}{d\lambda}
 \approx - \frac{1}{2}\, \partial^\mu (\phi^2)  \;, \qquad
\left(\frac{d\tau}{d\lambda}\right)^{\!2} \approx \phi^2 \qquad\qquad \mathrm{for\ \ } r \ll \kappa
\end{equation}
From (\ref{Psol}) follows that $\phi(r) \approx \sqrt{3} (r/\kappa)^{p-1/\sqrt{12}}$ for $r \ll \kappa$. Using this and (\ref{Zpgr0}) in (\ref{wgpe}) we obtain that the proper time of particle radially approaching $r=0$ behaves as $\tau \sim r^{1-p}$, which means that $\tau \to \infty$ for $r \to 0$ so the worldlines of particles are not terminated at $r=0$ at finite proper time. Altogether, we obtain that the surface $r=0$ is located "at infinity" so that the space-time is in a sense complete.

\medskip
\noindent
{\bf (b) Importance of quantum gravity effects.}\\[1mm]
In $p$-gauges it is still expected for quantum effects to be dominant near $r=0$. This can be seen by observing that effective Planck mass in $p$-gauge is given by
\begin{equation} 
(M_{\mathrm{Pl\,eff}})^2 = \frac{1}{12} \left( \phi^2 - s^2 \right)
 = \left( 1 + \frac{\kappa}{r} \right)^{\!-2p} \approx \left( \frac{r}{\kappa} \right)^{\!2p} \;, \qquad r \ll \kappa
\end{equation}
It becomes arbitrarily small near $r=0$, signalling the dominance of quantum gravity fluctuations in this region.

\medskip
\noindent
{\bf (c) Regularity and observables.}\\[1mm]
As observables are normally gauge invariant objects, we must ascertain that there is a large enough set of diff- and 
Weyl-invariant quantities which are regular (i.e., finite when letting $r\to0$). It is easy to see that there is an infinite number of non-trivial (i.e., not everywhere vanishing) regular local scalars including, e.g., 
\begin{equation} \label{regobs}
\frac{s}{\phi} \;,\qquad 1 - \frac{s^2}{\phi^2} \;,\qquad
 \phi^{-4} \left( 1 - \frac{s^2}{\phi^2} \right)^{\!5} C_{\mu\nu\rho\sigma}\, C^{\mu\nu\rho\sigma} \;,
\end{equation}
where $C_{\mu\nu\rho\sigma}$ is Weyl tensor, and all products of these scalars. In fact, one can take any Weyl-invariant scalar field and obtain from it (an infinite tower of) regular Weyl-invariant scalars by multiplying with $(1-s^2/\phi^2)^k$ with large enough exponent $k$ (and possibly other regular Weyl-invariant scalars).\footnote{Of course, not all scalars are expected to be regular on all physical configurations. For example, in general relativity with minimally coupled scalar field 
$\phi$ (especially if a global minimum of the potential is at $\phi=0$) such ``singular'' scalars are $\phi^{-k}$, $\phi^{-k} R$, etc., with $k>0$. We emphasise this trivial fact because of an objection, originally used in \cite{Carrasco:2013hua} in a cosmological context, that a Weyl-invariant $(\phi^2 - s^2)^{-2} C^2$ is singular when $r\to0$. However, it is not clear why this invariant should have such physical significance in WIDG gravity. On the contrary, from its form it is quite possible that it belongs to the class of singular invariants which have the problem of dividing-by-zero.} In a similar fashion we can also construct non-local global Weyl-invariant scalars using (single or multiple) integrals over spacetime which are regular for 
$r\to0$ (e.g., both kinetic and potential part of the WIDG action are such).

One important observable which must be addressed is the ``physical proper time'', which obviously is not the proper time defined from the metric due to its Weyl-nonivariance. The solution can be found in the equation (\ref{Aparticle}), which must be taken as a definition of ``the physical proper time''. Now, if we take $\chi(x)\to\textrm{const}$ as $x\to1$, then it is easy to show that the surface $r=0$ is separated by the finite physical time from the points with $r\ne0$. In this case it is possible to extend the solution, in the essentially smooth way\footnote{Regular Weyl-invariants behave smoothly over $r=0$.}, to the second asymptotic region of normal gravity or even to the anti-gravity regions of the parameter space in the similar manner as it was done in the cosmological setting in \cite{Bars:2011mh,Bars:2011aa,Bars:2012mt}. The second possibility happens when the function $\chi$ is of the form
\begin{equation}
\chi(s/\phi) = \left(1 - \frac{s^2}{\phi^2}\right)^{\!-k}
\end{equation}
For large enough $k$ physical time between $r=0$ and all $r\ne0$ points becomes infinite. In this case we can say that spacetime with $r>0$ is complete.

Which of the two possibilities is realised depends on the details of the physics near $r=0$. As argued above in (b), in the region $r \ll \kappa$ we expect full quantum gravity regime of some sort to be in operation so the proper analysis and behaviour of observables is impossible at the moment (though they are expected to be non-local, see, e.g., \cite{Giddings:2005id}). What we have shown is that in the (semi)classical sense there is no problem in constructing candidates for classical observables which are regular for $r\to0$, which is at least a promising starting point toward the full quantum description.

\medskip
\noindent
{\bf (d) Baby universe vs. white hole.}\\[1mm]
The interior of the BH solution in $p$-gauges has a sort of \emph{baby universe} type of metric. To see this, let us analyse two radial geodesics in angular coordinates $\theta$ and $\varphi$. The proper distance between geodesics as the function of time behaves as 
\begin{equation} 
D(r) \propto r^{\frac{1}{2}-p} \left( r + \kappa \right)^{\frac{1}{2}+p}
\end{equation}
The function $D(r)$ has a minimum at $r_{\mathrm{m}} = (p - 1/2) \kappa$, and tends to infinity when $r \to 0$ and 
$r \to \infty$. We see that radially infalling shell after the time $r_{\mathrm{m}}$ (i.e., for $r < r_{\mathrm{m}}$) expand its size. The baby universe is homogeneous but anisotropic. Now, taking into account not geometric but physical distances, this baby universe picture is unchanged in the second possibility stated in (c). In the first possibility it is possible to pass $r=0$ and go into other asymptotic regions, which resembles to the \emph{white hole} scenario.

\medskip
\noindent
{\bf (e) E-gauge singularity as a marker for the phase transition.}\\[1mm]
In $p$-gauges we have obtained that $\phi$, $s$ and $g^{\mu\nu}$ all vanish when $r \to 0$. It can be shown that the same applies to all fields with positive scaling dimension on Weyl-rescaling, such as, e.g., Weyl tensor $C^{\mu\nu}{}_{\rho\sigma}$.\footnote{This is valid for all definitions of time coordinate $\bar{t}=\bar{t}(r)$ in which surface $r=0$ is located at finite $\bar{t}_0 = \bar{t}(0)$. For other choices of time variable one gets that pullback of tensors on the surface $r=0$ is vanishing.} As $(\phi,s,g^{\mu\nu}) = (0,0,0)$ is a fixed point of Weyl transformations, we see that at a ``singularity'' $r=0$ there is a phase transition from broken to unbroken phase of Weyl symmetry. 

\begin{figure}[htb]
\begin{center}
\includegraphics[scale=0.8]{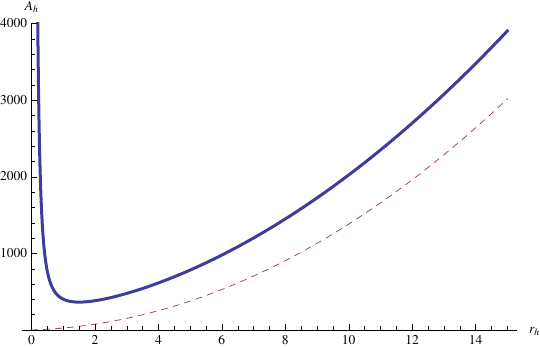}
\caption{Figure shows horizon areas $A_h$ in $p$-gauge (bold line) and in E-gauge (dashed line) as a function of coordinate horizon radius $r_h$, for parameters $p=2$ and $\kappa=1$. The mass $M$ of the BH is gauge independent, and $M/\kappa$ is a monotonic function of $r_h/\kappa$ (defined implicitly by $N(r_h)=0$). The exact profile of $A_h$ is gauge-dependent, but the divergence as $r_h \to 0$ is a generic feature in all regular gauges.}
\label{fig:area}
\end{center}
\end{figure}

\medskip
\noindent
{\bf (f) Thermodynamical properties and the fate of an isolated BH.}\\[1mm]
The BH entropy (as given by Wald formula), BH temperature (obtained from surface gravity which is conformally invariant) and ADM mass (defined by conformally invariant formula \cite{Chan:1996sx}) are invariant on Weyl transformations (\ref{weyltr}), so in $p$-gauges they are the same functions of parameters $\lambda$ and $\kappa$ as when calculated in E-gauge (i.e., in GR description). However, as the metric itself is not gauge-invariant, there are some new moments when one passes to regular gauges like $p$-gauges. Above we have showed this by analysing proper time and distance inside BH. Another example is the horizon area, which in $p$-gauge is different then in E-gauge, and is given by
\begin{equation}
A_h^{(p)} = \left( 1 + \frac{\kappa}{r_h} \right)^{\!2p} A_h^{(\mathrm{E})}
 = 4\pi\, r_h^2 \left( 1 + \frac{\kappa}{r_h} \right)^{\!2p+1}
\end{equation}
For large BHs ($r_h \gg \ell_{\mathrm{Pl}}$) we saw that $r_h \gg \kappa$ and 
$A_{(p)h} \approx A_{(\mathrm{E})h}$. In fact, for large BH in $r > r_h$ region solution in $p$-gauge is approximately equal to solution in E-gauge, which is approximately equal to Schwarzschild solution. However, as black hole becomes smaller, e.g., by Hawking radiation, and $r_h$ becomes of the order of $\kappa$, the factor $(1+\kappa/r)^{2p}$ becomes more and more important. A difference is dramatic for small black holes with $r_h \ll \kappa$ for which $A_{(p)h} \approx \kappa^2 (\kappa/r)^{2p-1}$, which diverges in the limit $r_h \to 0$. In Fig.\ \ref{fig:area} we present plots for horizon area as a function of horizon radius, both in $p$-gauge (with $p=2$) and $E$-gauge. The function $A_{(p)h}(r_h)$ has a global minimum at $r_{hm} = (p - 1/2) \kappa$, and such ``minimal" BHs are characterised by $\lambda\kappa_{\mathrm{m}}^2 \approx 3p$. Again, in the baby universe scenario the above analysis remains qualitatively the same after geometric proper distances are substituted with physical proper distances.

The properties of BH solution in $p$-gauges give a new twist to the ongoing debate on the question of what happens in the final phase of evaporation of BHs, which is crucially important for discussions going around information loss paradox. As an isolated BH, by emitting a Hawking radiation, reaches the final phase of evaporation the question is does it evaporates completely or there remains some sort of a remnant (baby universe, Planck-size quantum remnant, etc.). As the resolution depends on a yet not enough understood properties of the quantum gravity, this is an open question. Yet, it is frequently claimed that the option with remnant is not valid, or highly improbable. Standard arguments typically use one (or both) of the following properties of BH in GR: (1) ``small internal volume" (consequence of the classical singularity), (2) ``small size" as viewed from the outside (a consequence that classically $A_{(\mathrm{E})h} \to 0$ as $r_h \to 0$). However, we have shown that in regular gauges, like $p$-gauges, neither of these properties are present classically, first because there may be a baby universe (or a white hole) inside the BH, and second because the horizon ``explodes'' in the final stage of evaporation (i.e., $A_{(p)h} \to \infty$ for $r_h \to 0$) in the case of the baby universe scenario. Of course, one still needs a knowledge of the quantum gravity to fully answer the question what finally happens with the black hole, but in WIDG theories it appears that possibility of some sort of a BH remnant (or white hole) cannot be so easily dismissed. 

\medskip
\noindent
{\bf (g) Generic behaviour in regular gauges.}\\[1mm]
We have seen that $|s(x)/\phi(x)| \to 1$ when $r \to 0$, so that E-gauge breaks when singularity is approached and one needs to use another gauge which is regular in this limit. We have been able to construct a class of gauges, $p$-gauges (\ref{pgauge}), which are regular in the sense that all degrees of freedom ($\{\phi,s,g^{\mu\nu}\}$) are regular everywhere. It can be shown that while details depend on the choice of the gauge, the generic features described in (a)-(f) are gauge-independent in the sense that they are valid in all regular gauges. We used $p$-gauges just for demonstration purposes. Let us note that there may be other constraints on gauges to be regular, coming from the perturbation analysis (we discus this in Sec. \ref{sec:general}), quantisation and/or asymptotic behaviour at $r \to 0$.

On the example of BH studied above, we can understand why the new properties discovered here in $p$-gauge (and valid in all regular gauges), were not found before. E-gauge (\ref{Egauge}) and c-gauge (defined by $\phi(x)=\phi_0$) do not cover $(\phi,s) = (0,0)$ point, which makes them inappropriate for studying BH solution near $r = 0$ and this explains presence of singularity in these gauges. It is easy to show that in s-gauge (obtained by putting $p=1/2$ in (\ref{pgauge})) the metric of this BH solution is also singular at $r=0$. The classical equations in the four mentioned gauges have the form of standard GR (minimally coupled fields), non-minimal GR with conformally coupled fields, and GR in ``string-frame''. These are the most studied formulations of gravity with 2-derivative actions in the context of constructing classical solutions, and, as we see, they all correspond to singular gauges.


\section{Generalisation}
\label{sec:general}


One can check that the analysis from the previous section can be repeated, with essentially the same conclusions, for \emph{any of the known spherically symmetric BH solutions in GR with non-trivial scalar field, regardless is it asymptotically flat or AdS, as long as they describe proper BH's from the viewpoint of the WIDG formulation of gravity}. In particular this is true for charged dilatonic BH solutions in low energy superstring effective theories and string inspired theories reviewed in \cite{Horowitz:1992jp,Youm:1997hw}, or for BHs with non-Abelian hair reviewed in 
\cite{Lavrelashvili:1997yf}, which all have stable classical vacuum. The key property is that the scalar field in all these solutions is of the form $\sigma(r) \approx C_1 \ln(r-r_s) + C_2$ near the singularity $r=r_s$ (in "Schwarzschild-like" coordinates), which then enables one to construct gauges, e.g., $p$-gauges (\ref{pgauge}) with $p$ larger then some $p_0$, in which the solution is regular everywhere and analysis from the previous section can be readily performed. 

For example, let us take as our second example the string inspired theory containing the scalar field $\sigma$ and $U(1)$ gauge field $A_\mu$, with the Lagrangian given by
\begin{equation}
\mathcal{L}_E = R_E - \frac{1}{2} (\partial_\mu \sigma)^2 - e^{-\sigma} (F_{\mu\nu})^2
\end{equation}
where $F_{\mu\nu} = \partial_\mu A_\nu - \partial_\nu A_\mu$, and again using the convention $16\pi G_N = 1$. Contrary to the example used in the previous section, this theory does not have the problem with the classical stability of the vacuum. It was shown in \cite{Gibbons:1982ih} that there is a classical solution given by
\begin{eqnarray}
&& ds_E^2 = - \left( 1 - \frac{2M}{r} \right)\, dt^2 + \left( 1 - \frac{2M}{r} \right)^{-1}\, dr^2
 + r \left( r - \frac{Q^2}{M} \right)\, d\Omega_2
\nonumber \\
&& \sigma(r) = \ln \left( 1 - \frac{Q^2}{M r} \right) \qquad\qquad,\qquad\qquad F_{rt} = \frac{Q}{r^2}
\end{eqnarray}
This solution describes electrically charged asymptotically flat BH where $M$ is the mass and $Q$ is the electric charge $Q$. Event horizon is located at $r_h = 2M$ and singularity at $r_s = Q^2/M$, so the radial coordinate in E-gauge is defined only for $r > Q^2/M$. We can closely follow the procedure from the previous section, show that in $p$-gauges (\ref{pgauge}), with $p > 1$, solution is regular in the sense described there, and obtain the same general conclusions as in (a)-(g).

However, our construction obviously does not work for BH solutions in which scalar field is trivial, i.e., constant in the whole space-time, which include the very important case of Schwarzschild BHs. Moreover, for a large class of scalar potentials $V_{\mathrm{E}}(\sigma)$, there are ``no-hair'' theorems guarantying that Schwarzschild BH with constant scalar field is the unique spherically symmetric BH solution regular everywhere except at a central singularity.\footnote{In the example from the previous section the potential (\ref{Zpot}) is unbounded from below, which is the reason why ``no-hair'' theorems are avoided.} It is an obvious consequence of (\ref{pgauge}) that in this case solution in $p$-gauge has the same form as in E-gauge so it is singular at $r=0$. Constant  $\sigma$ means, by (\ref{phigE}), that both ``fundamental''  scalar fields $\phi_{\mathrm{E}}$ and $s_{\mathrm{E}}$ are constant and moreover that 
$|s(x)| < |\phi(x)|$ everywhere including the singularity $r=0$. We see that Schwarzschild BH violates the central conjecture that GR singularities are artefacts of invalidity of E-gauge when $|s| = |\phi|$. Does this kill our argument? No. The clue lies in understanding that Schwarzschild BH is a singular solution in the space of all BH solutions which is realised only for idealised initial conditions (of the zero measure). When one considers perturbations around it, either coming from matter/fields infalling or sitting outside BH \cite{Doroshkevich:1978aq}, or when BH is inside the dynamical background (like FRW metric of cosmological setting \cite{Chadburn:2013mta}), the perturbed fields generally behave near the singularity $r=0$ as $\Phi(x) \sim \ln(r)$ for scalar fields and as $\Phi(x) \sim r^{-n}$ for higher-spin fields (in the "Schwarzschild-like" coordinates). We see that for generic solutions, corresponding to generic initial conditions, we have $\sigma \sim \ln(r) \to \infty$ which directly implies $|s/\phi| \to 1$ when $r\to0$. Let us mention that general non-perturbative analysis of space-like singularities confirms that logarithmic behaviour of scalar fields is generic \cite{BK1973}. We then conjecture that there are regular gauges in which \emph{generic} BH solutions are described by finite and regular $\phi(x)$, $s(x)$ and $g_{\mu\nu}(x)$ (and other matter fields) in the whole space-time.

The full procedure of finding well-defined gauges in which solutions are regular cannot be performed simply because there are no analytic solutions for such ``perturbed'' Schwarzschild BHs. To get a taste of what is happening, let us consider a related problem of regulating generic spherically symmetric homogeneous solutions near a space-like singularity in GR. In GR (E-gauge) with a massless scalar with vanishing potential $V_{\mathrm{E}}(\sigma) = 0$ such generic solutions near space-time singularity, located at $r=0$, are approximately of the form \cite{Burko:1997xa}
\begin{equation} \label{GRsingh}
ds_{\mathrm{E}}^2 \approx r^{\beta^2-1} dt^2 - A\, r^{\beta^2+1} dr^2
 + r^2 (d\theta^2 + \sin^2\theta d\varphi^2) \;, \qquad
 \sigma \approx \beta \ln r + C
\end{equation}
where $C$, $\beta$ and $A>0$ are independent constants of integration, and $r=0$ defines space-time singularity. The singularity of Schwarzschild BH with mass $M$ is obtained by setting $\beta = 0$ and $A = 1/(2GM)$. We explicitly see the exceptional character of Schwarzschild BH singularity.\footnote{We also see that the BH solution (\ref{Zmetric})-(\ref{Zsoldef}), which we used as an example to demonstrate our claims, has a singularity which is of generic type.} For all other solutions one has $\beta \ne 0$ and logarithmic behaviour of the field $\sigma$. It is easy to find gauges in which solutions with $\beta\ne0$ are regular, e.g., by choosing $\Omega = \exp(-\sigma^2)$ as a Weyl-rescaling factor doing transition from E-gauge.\footnote{The $p$-gauges we constructed in Sec. \ref{sec:example} are not completely satisfactory here because for fixed finite $p$ they would regulate only solutions (\ref{GRsingh}) having $\beta \ge \beta_p$, where $\beta_p>0$ is a function of $p$.} In all such gauges the same picture emerges as in the example studied in Sec. \ref{sec:example}, again confirming generic nature of our results. 

\begin{figure}[htb]
\begin{center}
\includegraphics[scale=0.8]{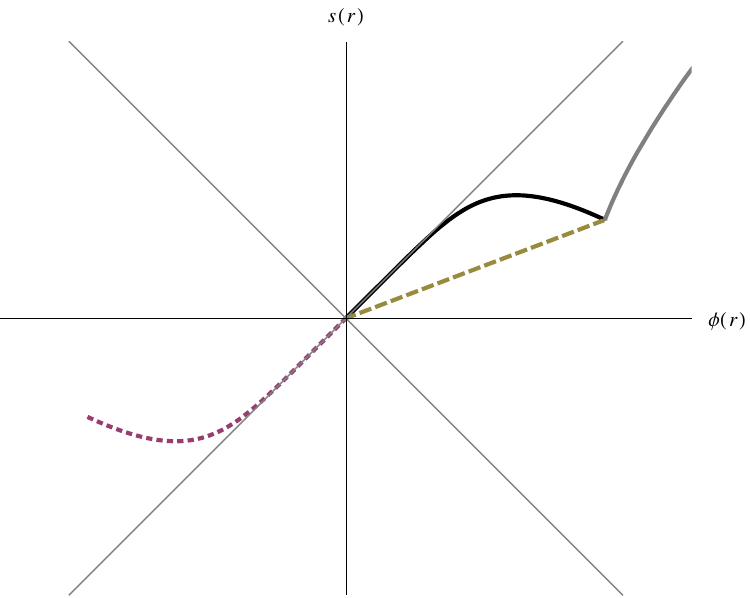}
\caption{The figure shows a non-generic nature of Schwarzschild singularity, and also a singular nature of E-gauge (GR description) by plotting solutions in $(\phi,s)$-plane. The black thick line represents a BH solution with generic spacelike (E-gauge) singularity which is in regular gauges characterised by $(\phi,s) \to (0,0)$ and $|s/\phi| \to 1$ (independently of values of scalar fields in asymptotic infinity). The dashed line represents Schwarzschild BH which is in all gauges characterised by $|s/\phi| = \mathrm{const} < 1$. The point in which two solutions coincide corresponds to asymptotic infinity $r\to\infty$. Solutions in E-gauge are obtained by projecting on a hyperbola (thin gray line). We see that for generic solution this projection (thick gray line) is ill-defined when $(\phi,s) \to (0,0)$ in regular gauge, and this manifests as space-time singularity in E-gauge. The dotted thick line represents a white hole extension (in region with negative $\phi$ and $s$). Thin grey diagonal lines are straight lines $|\phi| = |s|$.}
\label{fig:gensol}
\end{center}
\end{figure}

There are gauges in which Schwarzschild BH is regular, obtained by using higher-derivative terms. An example, with lowest-order in derivatives, is a class of gauges obtained by applying Weyl-factor
\begin{displaymath}
\Omega = \left(1 + c \, (R_{\mu\nu\rho\sigma})^4 \right)^{-p} \;, \qquad c>0 \;, \qquad p > 1/8
\end{displaymath}
on the E-gauge fields.\footnote{In these gauges classical action contains higher-derivative terms typically leading to new degrees of freedom and/or breaking of unitarity. For this reason results obtained in such gauges should be treated with care.} However, in these gauges Schwarzschild solution is still exceptional because it has $|s/\phi| = \mathrm{const} < 1$ (this constant is determined by the asymptotic value at $r \to \infty$) while generic solutions behave as $|s/\phi| \to 1$ in $r\to0$ limit. This is pictorially described in Fig.\ \ref{fig:gensol}. Beside, in such gauges one introduces higher-derivative terms which are usually unwanted (though they should't be dismissed a priori).

The emerging picture from our analysis is that \emph{Schwarzschild BH is a singular solution also in WIDG formulation, and should not be considered ``microscopically'' relevant} (it can be used only as an average over ensemble of regular classical solutions). This means that it may be misleading to use calculations based only on Schwarzschild BH as a guideline in circumstances where microscopic properties of the near-singularity region play an important role. This is not so surprising as the same is probably true for other eternal BH solutions, like Reissner-Nordstr\"{o}m or Kerr BHs, which poses inner Cauchy horizons which appear to be unstable under small perturbations. The morale is: To understand what is happening inside the BHs, one must consider solutions with initial conditions which are generic enough.


\section{Conclusion}
\label{sec:concl}


We have demonstrated that a two-derivative Weyl-invariant formulation of gravity provides a description of black holes essentially without classical singularities. Appearance of singularities in standard formulation of gravity (GR) is simply an artefact of the fact that particular choice of the gauge fixing of local scale symmetry, in which the theory is equivalent to GR, is not well-defined at singularities. By studying classical black hole solutions in regular gauges, in which fundamental fields are all well-defined in the whole space-time, we obtained the following main results: (1) singularity is a signal of a phase transition from broken to unbroken phase of Weyl symmetry,  (2) instead of a singularity there is a ``baby-universe'' or white hole inside a black hole, (3) as black hole decreases its mass (e.g., by Hawking radiation) there is a critical mass after which reducing mass makes black hole bigger as viewed by outside observers (in the baby universe scenario). We emphasise that these results are obtained in the theory which is classically a completion of GR, without introducing new physical fields (normal or phantom) or higher-derivative terms in the action. In fact, our classical analysis is essentially equivalent to using particular field redefinitions, but allowing the possible natural extension of the parameter space (important in white hole scenario). The essential requirement is that there is at least one physical scalar field in the theory with the properties which Standard Model Higgs field fulfils.

These results may be of relevance for issues related to unitarity of the evolution of black holes, indicating that BHs do not evaporate completely through the process of Hawking radiation and that some information, of possibly unlimited amount, may simply go into remnant baby universe or white hole. Such scenarios may offer a resolution (see, e.g., 
\cite{Hsu:2006pe,Hossenfelder:2009xq} for a review) of Hawking's information loss paradox without resorting to exotic concepts, such as ``firewalls'', breaking of semi-classical approximation or violation of unitarity. 

It is interesting to compare our findings with those obtained in the context of cosmological singularities inside the same framework of Weyl-invariant formulation of gravity and presented in \cite{Bars:2011aa,Bars:2012mt}. The common observation is that generic solutions in regular gauges behave as $(\phi,s) \to (0,0)$ and $|s/\phi| \to 1$ as one approaches surfaces which are homogeneous spacelike singularities in E-gauge, while the simple solutions, like highly symmetric FRW solutions (homogeneous and isotropic) and Schwarzschild BH (constant scalar fields), have different non-generic behaviour. Also, any attempt to extend solutions through the GR singularity necessarily implies breakdown of classical spacetime metric description, but there is enough number of classical observables which are regular across the singularity which can be used to formally make such extensions.\footnote{After the first version of this paper was submitted, a paper \cite{Carrasco:2013hua} which emphasised singular behaviour of extensions through cosmological singularities constructed in \cite{Bars:2011aa,Bars:2012mt}. However, as explained in point (c) in Sec. \ref{sec:example} (see in particular footnote 3), we think that the particular argument used in \cite{Carrasco:2013hua} is too simplistic.} These agreements indicate that our main conclusions may possibly generalise to all spacelike singularities. 

The analysis we presented is completely classical, and for several reasons it cannot be considered complete without mentioning quantum effects. First, one could worry about health of the Weyl invariance we used extensively, in view of the generally anomalous nature of scale symmetry. However, in theories with dilaton field there are regularisations in which classical Weyl invariance is preserved in an effective action \cite{Percacci:2011uf,Codello:2012sn}. Second, and indeed very important, reason is that the proposed classical resolution of singularities happens deeply in the quantum regime, so knowledge of the quantum gravity is necessary to know can we trust our classical picture of regular spacetime (or picture with extension to white hole) or not. Needless to say, such knowledge is currently beyond our grasp. But, building the quantum theory around a regular classical solutions (situation in Weyl-invariant formulation of gravity), as opposed to a singular solutions (a situation in standard General Relativity), would normally be considered as a much better starting point.

\vspace{20pt}

\noindent
{\bf \large Acknowledgements}

\bigskip

\noindent
We thank L. Bonora, M. Cvitan, S. Pallua and I. Smoli\'{c} for stimulating discussions. The research was supported by University of Rijeka under the research support No. 13.12.1.4.05. The author declares that there is no conflict of interest regarding the publication of this paper.




\begin{thebibliography}{99}


\bibitem{Kanti:1998jd}
  P.~Kanti, J.~Rizos and K.~Tamvakis,
  ``Singularity free cosmological solutions in quadratic gravity,''
  Phys.\ Rev.\ D {\bf 59}, 083512 (1999) \ 
  [gr-qc/9806085].

\bibitem{Modesto:2011kw}
  L.~Modesto,
  ``Super-renormalizable Quantum Gravity,''
  Phys.\ Rev.\ D {\bf 86}, 044005 (2012) \ 
  [arXiv:1107.2403 [hep-th]].

\bibitem{Biswas:2011ar}
  T.~Biswas, E.~Gerwick, T.~Koivisto and A.~Mazumdar,
  ``Towards singularity and ghost free theories of gravity,''
  Phys.\ Rev.\ Lett.\  {\bf 108}, 031101 (2012) \
  [arXiv:1110.5249 [gr-qc]].

\bibitem{Bars:2013yba}
  I.~Bars, P.~Steinhardt and N.~Turok,
  ``Local Conformal Symmetry in Physics and Cosmology,''
  arXiv:1307.1848 [hep-th].

\bibitem{Dirac:1973gk}
  P.~A.~M.~Dirac,
  ``Long range forces and broken symmetries,''
  Proc.\ Roy.\ Soc.\ Lond.\ A {\bf 333}, 403 (1973).

\bibitem{Kallosh:2013hoa}
  R.~Kallosh and A.~Linde,
  ``Universality Class in Conformal Inflation,''
  JCAP {\bf 1307}, 002 (2013) \
  [arXiv:1306.5220 [hep-th]].

\bibitem{Hooft:2010ac}
  G.~'t Hooft,
  ``Probing the small distance structure of canonical quantum gravity using the conformal group,''
  arXiv:1009.0669 [gr-qc]; 

\bibitem{tHooft:2011aa}
  G.~'t Hooft,
  ``A class of elementary particle models without any adjustable real parameters,''
  Found.\ Phys.\  {\bf 41}, 1829 (2011) \
  [arXiv:1104.4543 [gr-qc]].

\bibitem{Bars:2011mh}
  I.~Bars, S.~-H.~Chen and N.~Turok,
  ``Geodesically Complete Analytic Solutions for a Cyclic Universe,''
  Phys.\ Rev.\ D {\bf 84}, 083513 (2011) \
  [arXiv:1105.3606 [hep-th]].

\bibitem{Bars:2011aa}
  I.~Bars, S.~-H.~Chen, P.~J.~Steinhardt and N.~Turok,
  ``Antigravity and the Big Crunch/Big Bang Transition,''
  Phys.\ Lett.\ B {\bf 715}, 278 (2012) \ 
  [arXiv:1112.2470 [hep-th]].

\bibitem{Bars:2012mt}
  I.~Bars, S.~-H.~Chen, P.~J.~Steinhardt and N.~Turok,
  ``Complete Set of Homogeneous Isotropic Analytic Solutions in Scalar-Tensor Cosmology with  
  Radiation and Curvature,''
  Phys.\ Rev.\ D {\bf 86}, 083542 (2012) \ 
  [arXiv:1207.1940 [hep-th]].

\bibitem{Fisher48}
  I.~Z.~Fisher,
  ``Scalar mesostatic field with regard for gravitational effects'',
  Zh.\ Eksp.\ Teor.\ Fiz.\ {\bf 18}, 636 (1948) \
  [arXiv:gr-qc/9911008].

\bibitem{JNW68}
  A.~I.~Janis, E.~T.~Newman and J.~Winicour,
  ``Reality of the Schwarzschild singularity'',
  Phys.\ Rev.\ Lett.\  {\bf 20}, 878 (1968).

\bibitem{Zloshchastiev:2004ny}
  K.~G.~Zloshchastiev,
  ``On co-existence of black holes and scalar field,''
  Phys.\ Rev.\ Lett.\  {\bf 94}, 121101 (2005) \ 
  [hep-th/0408163].

\bibitem{BBM}
  N.~Bocharova, K.~Bronnikov and V.~Melnikov,
  Vestn.\ Mosk.\ Univ.\ Fiz.\ Astron. {\bf 6}, 706 (1970).

\bibitem{Prester:2014voa}
  P.~Dominis Prester,
  ``Field redefinitions, Weyl invariance and the nature of mavericks,''
  Class.\ Quant.\ Grav.\  {\bf 31}, 155006 (2014) \
  [arXiv:1405.1941 [gr-qc]].

\bibitem{Bekenstein:1974sf}
  J.~D.~Bekenstein,
  ``Exact solutions of Einstein conformal scalar equations,''
  Annals Phys.\  {\bf 82}, 535 (1974).
 
\bibitem{Martinez:2002ru}
  C.~Martinez, R.~Troncoso and J.~Zanelli,
  ``De Sitter black hole with a conformally coupled scalar field in four-dimensions,''
  Phys.\ Rev.\ D {\bf 67}, 024008 (2003) \
  [hep-th/0205319].

\bibitem{Cebeci:2000pt}
  H.~Cebeci and T.~Dereli,
  ``Conformal black hole solutions of axi dilaton gravity in D-dimensions,''
  Phys.\ Rev.\ D {\bf 65}, 047501 (2002) \
  [gr-qc/0009069].
  
\bibitem{Bekenstein:1975ts}
  J.~D.~Bekenstein,
  ``Black Holes with Scalar Charge,''
  Annals Phys.\  {\bf 91}, 75 (1975).

\bibitem{Carrasco:2013hua}
  J.~J.~M.~Carrasco, W.~Chemissany and R.~Kallosh,
  ``Journeys Through Antigravity?,''
  JHEP {\bf 1401}, 130 (2014) \
  [arXiv:1311.3671 [hep-th]].

\bibitem{Giddings:2005id}
  S.~B.~Giddings, D.~Marolf and J.~B.~Hartle,
  ``Observables in effective gravity,''
  Phys.\ Rev.\ D {\bf 74}, 064018 (2006) \ 
  [hep-th/0512200].
  
\bibitem{Chan:1996sx}
  K.~C.~K.~Chan, J.~D.~E.~Creighton and R.~B.~Mann,
  ``Conserved masses in GHS Einstein and string black holes and consistent thermodynamics,''
  Phys.\ Rev.\ D {\bf 54}, 3892 (1996) \
  [gr-qc/9604055].

\bibitem{Anabalon:2013qua}
  A.~Anabalon, D.~Astefanesei and R.~Mann,
  ``Exact asymptotically flat charged hairy black holes with a dilaton potential,''
  JHEP {\bf 1310}, 184 (2013) \ 
  [arXiv:1308.1693 [hep-th]].

\bibitem{Anabalon:2012dw}
  A.~Anabalon,
  ``Exact Hairy Black Holes,''
  arXiv:1211.2765 [gr-qc].

\bibitem{Horowitz:1992jp}
  G.~T.~Horowitz,
  ``The dark side of string theory: Black holes and black strings.,''
  In *Trieste 1992, Proceedings, String theory and quantum gravity '92* 55-99
  [hep-th/9210119].

\bibitem{Youm:1997hw}
  D.~Youm,
  ``Black holes and solitons in string theory,''
  Phys.\ Rept.\  {\bf 316}, 1 (1999)
  [hep-th/9710046].

\bibitem{Lavrelashvili:1997yf}
  G.~V.~Lavrelashvili,
  ``NonAbelian surprises in gravity,''
  gr-qc/9701049.

\bibitem{Gibbons:1982ih}
  G.~W.~Gibbons,
  ``Antigravitating Black Hole Solitons with Scalar Hair in N=4 Supergravity,''
  Nucl.\ Phys.\ B {\bf 207}, 337 (1982).

\bibitem{Doroshkevich:1978aq}
  A.~G.~Doroshkevich and I.~D.~Novikov,
  ``Space-Time and Physical Fields in Black Holes,''
  Zh.\ Eksp.\ Teor.\ Fiz.\  {\bf 74}, 3 (1978).

\bibitem{Chadburn:2013mta}
  S.~Chadburn and R.~Gregory,
  ``Time dependent black holes and scalar hair,''
  arXiv:1304.6287 [gr-qc].
  
\bibitem{BK1973}
  V.~A.~Belinskii and I.~M.~Khalatnikov,
  ``Effect of scalar and vector fields on the nature of the cosmological singularity'',
  Zh.\ Eksp.\ Teor.\ Fiz.\ {\bf 63}, 1121 (1972) [Sov.\ Phys.\ JETP {\bf 36}, 591 (1973)].

\bibitem{Burko:1997xa}
  L.~M.~Burko,
  ``Homogeneous space - like singularities inside spherical black holes,''
  In *Haifa 1997, Internal structure of black holes and spacetime singularities* 212-233 \ 
  [gr-qc/9711012].

\bibitem{Hsu:2006pe}
  S.~D.~H.~Hsu,
  ``Spacetime topology change and black hole information,''
  Phys.\ Lett.\ B {\bf 644}, 67 (2007) \
  [hep-th/0608175].

\bibitem{Hossenfelder:2009xq}
  S.~Hossenfelder and L.~Smolin,
  ``Conservative solutions to the black hole information problem,''
  Phys.\ Rev.\ D {\bf 81}, 064009 (2010) \
  [arXiv:0901.3156 [gr-qc]].

\bibitem{Percacci:2011uf}
  R.~Percacci,
  ``Renormalization group flow of Weyl invariant dilaton gravity,''
  New J.\ Phys.\  {\bf 13}, 125013 (2011) \ 
  [arXiv:1110.6758 [hep-th]].

\bibitem{Codello:2012sn}
  A.~Codello, G.~D'Odorico, C.~Pagani and R.~Percacci,
  ``The Renormalization Group and Weyl-invariance,''
  Class.\ Quant.\ Grav.\  {\bf 30}, 115015 (2013) \
  [arXiv:1210.3284 [hep-th]].


\end{thebibliography}
\end{document}